# Lessons from Oz: Design Guidelines for Automotive Conversational User Interfaces

**David R. Large**
Human Factors Research Group,
University of Nottingham,
Nottingham, UK
david.r.large@nottingham.ac.uk

**Gary Burnett**
Human Factors Research Group,
University of Nottingham,
Nottingham, UK
gary.burnett@nottingham.ac.uk

**Leigh Clark**
School of Information and
Communication Studies,
University College Dublin,
Dublin, Ireland
leigh.clark@ucd.ie

## Abstract
This paper draws from literature and our experience of conducting Wizard-of-Oz (WoZ) studies using natural language, conversational user interfaces (CUIs) in the automotive domain. These studies have revealed positive effects of using in-vehicle CUIs on issues such as: cognitive demand/workload, passive task-related fatigue, trust, acceptance and environmental engagement. A nascent set of human-centred design guidelines that have emerged is presented. These are based on the analysis of users' behaviour and the positive benefits observed, and aim to make interactions with an in-vehicle agent interlocutor safe, effective, engaging and enjoyable, while conforming with users' expectations. The guidelines can be used to inform the design of future in-vehicle CUIs or applied experimentally using WoZ methodology, and will be evaluated and refined in ongoing work.

## CCS Concepts
• **Human-centered computing~Natural language interfaces**

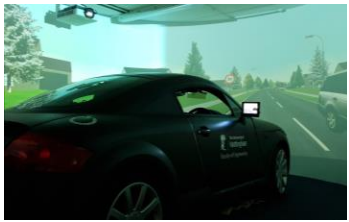
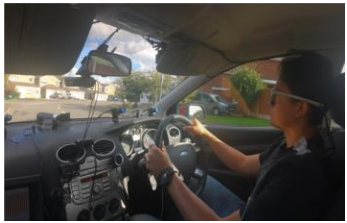
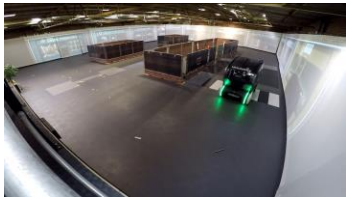

Figure 1: Wizard-of-Oz studies have taken place in driving simulators (top), on-the-road and in autonomous, self-driving pods (bottom).

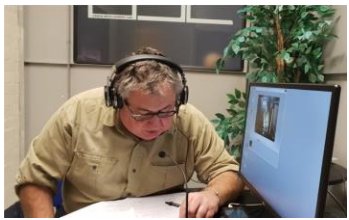

Figure 2: Professional actor assuming role of 'Wizard'.

## Introduction

Speech-based, natural language interactions (NLI) enabled by so-called conversational user interfaces (CUI) are increasingly prevalent in a number of contexts, such as the home, personal mobile devices and driving, and are expected to become the *de facto* method of interaction in future technologies. This is primarily because CUI offer a quick, intuitive and increasingly reliable means of interaction: in theory, they do not require users to learn a new interaction technique, but instead rely on the use of a natural and familiar approach, namely speaking. Consequently, there appears to be a strong desire to emulate human-human (H-H) conversation (in all its subtleties) within voice interaction designs, in some cases including the addition of bogus fillers (*um, err, uh* etc.) [1]. Yet, evidence suggests that users may not expect or perceive the interaction to be conversation per se, but rather one that enables conversational exchanges [2], and more typically choose command-based (call-and-response) utterances [3, 4], only engaging in more recognisable 'conversation' as a repair mechanism when things go wrong. Moreover, the principles of good user interface design are also applicable to voice interfaces, and therefore understanding the user, their goals and the specific context of use is paramount [5]: it is reasonable to expect that the characteristics of CUIs (and the so-called conversation they enable) may differ significantly from one context to another.

Employing a Wizard-of-Oz (WoZ) approach [6], whereby a human actor has performed the role of the talking technology, we have conducted a corpus of studies exploring the effects of natural language interactions in the automotive domain (Figure 1). This work has revealed that an on-board agent with whom the driver can interact freely using conversational language can minimise cognitive demand and workload [7]. In addition, it can be an effective counter-measure to passive task-related fatigue [8, 9], increase levels of trust and acceptance in autonomous vehicles [10, 11] and enrich environmental engagement [12]. A particular challenge in conducting this work has been to create and deliver an 'authentic' CUI experience, not least to ensure that we do not violate users' high expectations of future technology or indeed how they expect, or would choose to interact.

## Wizard-of-Oz

WoZ is a popular, well-established technique in experimental HCI research that has been used successfully to evaluate future design concepts as well as conduct user acceptance studies on finalised interface designs [6]. In a typical WoZ study, a human 'wizard' simulates the technology remotely (in this case, by delivering verbal prompts and responding to users' commands and utterances), in a manner such that the user believes that the behaviour is system generated.

Conducting CUI-WoZ testing therefore necessitates a strict protocol, including a predefined 'script' outlining a range of opening gambits and appropriate responses that can be delivered in real-time. However, it also allows complete freedom and flexibility, enabling the 'wizard' to deal with all and any unexpected responses and events, and thus requires the talents of a skilled performer – in our case, a professional actor (Figure 2). Working from a script (initially inspired by literature, but also developed iteratively during the course of our work), our wizard delivered utterances in a controlled fashion using a subtle computer inflexion, subtly deviating from pure human enunciation. In addition, they were instructed to respond to all driver requests and avoid any clinical, out-of-domain responses, such as "*Sorry. I don't understand*", other than in the event of technical problems or delays retrieving information (in which case, "*Searching database…*" was employed instead). Finally, subtle pauses were introduced between fragments of each utterance to simulate the system accessing information and reconstructing the spoken statement, for example, "*I have looked at your to-do list and see that you must <pause> buy milk <pause> on your way home.*" While aspects of this approach may be at odds with the ultimate goal of voice-technology developers (i.e. to recreate H-H conversation), our work has led us to believe that users still expect imperfections, in so far as it avoids the sense of eeriness and suspicion that may accompany technology if it is perceived as too human (the so-called, 'uncanny valley' effect [13]). Thus, our overall WoZ approach aimed to exceed current state-of-the-art CUI, while conforming with our understanding of users' expectations of future conversational technology, and thereby avoiding the perils of the uncanny valley.

## Recommendations and Guidelines

We present an empirically-derived set of human-centred design guidelines outlining the key characteristics and conversational abilities of an on-board, agent-based CUI. These have been derived from the analysis of the conversations that took place between agent and vehicle occupant during our studies [4, 14], and further explored during focus groups and interviews with study participants [11]. They have been implicit in the development and embodiment of our in-vehicle conversational agent, and have provided demonstrable benefits in the driving domain.

1. *The agent should have a name and self-reference using the first person.* Giving something a name acknowledges identity and engenders trust. It also conforms with users' expectations of a conversational partner, and reinforces gender and personality. In a driving context, the name can also be used by drivers to grab attention (or 'barge-in') during interactions. In such situations, the system should stop speaking and start listening. The use of "I" appears to be expected if there is any indication of humanness (in this case, a human voice). It can also enhance the human-agent relationship because the interface is perceived to be more like a person (but within recognisable limits).

2. *The agent should have a clearly defined role as an assistant, including a relevant personality and emotional tone.* This allows drivers to maintain agency and responsibility over decision-making, but delegate task execution to the system [11]. Adding personality and emotions can enhance this perception and aid task execution and efficiency. It also ensures that agent is seen as subordinate, but still something that is socially-enabled [14].

3. *The agent should provide an introduction at the start of the interaction and instigate conversation where appropriate.* This conforms with people's experiences of human-human interactions (HHI). It increases the sense of agency and intelligence, and grounds the interaction. It enables the driver to build expectations, supporting the development of an appropriate mental model. Instigating conversation can also help combat issues of passive task-related fatigue [8, 9].

4. *The agent should engage in ('functional') small-talk.* This enables the driver to calibrate trust in the

agent, but care should be taken to avoid distracting the driver with complex or emotionally-laden topics (or with 'chit-chat'). It also provides scope to incorporate personalisation and seek driver preferences [4], and plays a significant role in avoiding the agent being perceived as too human.

5. *The agent should use pronouns to differentiate ownership of tasks.* Differentiating pronouns helps to distinguish which activities are socially marked – those that are conducted between participant and system, in which trust is shared (e.g. *our* journey) – and those which are considered to be individual (i.e. owned by the participant, or system, only, e.g. *my* diary). It also contributes to the development of 'appropriate' trust in both the agent and, by association, the host vehicle [4].

6. *Task-based responses should be consistent.* This ensures efficient task execution (minimising potential distraction) and avoids misunderstandings or errors. For example, the agent should confirm their understanding of an utterance or request ("*OK*"), should explicitly identify if their response is not immediately forthcoming ("*Searching database…*"), and confirm closure ("*That is done!*").

7. *The agent should explain its actions and decisions, wherever possible.* This ensures transparency of decision-making, and helps the user build an accurate mental model. It can also increase trust, and help overcome issues of privacy and security of information [11].

8. *The agent should moderate dialogue style, delivery of information and the pace of conversation based on driver workload/distraction*. This avoids distracting the driver at times of high-workload, and ensures they understand the relevance of information.

9. *The agent should employ social etiquette (politeness, apology) when delivering utterances or responding to the driver.* This enhances trust and the overall affective experience [14]. In addition, taking the blame (for example, through apology) can diffuse a difficult situation. Nevertheless, the agent should not expect the driver to respond accordingly – they are not constrained by the same social norms and expectations that exist in HHI, and may respond unexpectantly, or not at all.

10. *A tangible source or entity should be provided to represent/embody the agent.* This helps to create a sense of agency, differentiates the agent from the vehicle itself, and provides the capacity to switch it off (*shut it up*). Care should be taken that this does not visually distract the driver. Appropriate design can further help to distinguish the agent from being perceived as too human.

## Conclusion

These guidelines represent what we believe are the key characteristics and conversational abilities of an on-board, agent-based CUI. They aim to ensure that journey experiences are engaging and enjoyable, and that the agent interlocutor is perceived as highly capable, but without appearing too human. The guidelines also provide scope to address context-specific concerns of driver workload, fatigue and distraction, as well potentially enhancing trust in the technology, enriching environmental engagement, and addressing drivers' privacy and security concerns. They will be further evaluated and refined in future work.

**Acknowledgements**

We would like to thank all of our collaborators, in particular Jaguar Land Rover, the multitude of researchers who have helped to collect and analyse data, and finally, our resident wizard, Pablo Raybould, without whom we would still be stuck in Oz.